\documentclass[a4paper,twocolumn,amsmath,amssymb,pra,footinbib,floatfix,superscriptaddress]{revtex4}

\usepackage[utf8]{inputenc}
\usepackage{lmodern}
\usepackage[T1]{fontenc}

\usepackage{mathtools}
\usepackage{float, graphicx, caption, color, amsmath}
\usepackage[hidelinks]{hyperref}
\usepackage[left=1.9cm,right=1.9cm,top=1.5cm,bottom=1.8cm]{geometry}
\usepackage{graphics}
\usepackage{appendix}
\usepackage{amsmath}
\usepackage[none]{hyphenat}

\begin{document}

\title{\textbf{General quantum Chinos games}} 

\author{Daniel Centeno and Germ\'an Sierra \\ 
Instituto de F\'isica Te\'orica UAM-CSIC, Universidad Aut\'onoma de Madrid, Cantoblanco, 
 Madrid, Spain.}

\begin{abstract}
The Chinos game is a non-cooperative game between players who try to guess the total sum of coins drawn collectively. Semiclassical and quantum versions of this game were proposed by F. Guinea and M. A. Mart\'{\i}n-Delgado, in \textit{J. Phys. A: Math. Gen. 36 L197 (2003)}, where the coins are replaced by a boson whose number occupancy is the aim of player's guesses. Here, we propose other versions of the Chinos game using a hard-core boson, one qubit and two qubits. 
In the latter case, using entangled states the second player has a stable winning strategy that becomes symmetric for non-entangled states. Finally, we use the IBM Quantum Experience to compute the basic quantities
involved in the two-qubit version of the game. 
\end{abstract}

\maketitle

\def\beq{\begin{equation}}
\def\eeq{\end{equation}}
\def\barray{\begin{eqnarray}}
\def\earray{\end{eqnarray}}

  \thispagestyle{empty}

\newpage
\setcounter{page}{1}
\pagestyle{plain}
\section{Introduction}

	Game theory is a field that has fascinated mathematicians since the early 20th century. The idea is to be able to interpret complex problems in many different fields as a game of one or more players who, using logical reasoning, try to optimize their strategies to obtain the highest possible profit.
			The pioneers who formalized game theory were John von Neumann and Oskar Morgenstern \cite{Neumann}, followed soon later by John Nash  \cite{Nash}. A fundamental concept is that of the Nash equilibrium, in which it is assumed that every player knows and chooses their best possible strategy knowing the strategies of the other players. This implies that the Nash equilibrium situation is one in which no player would profit by changing her/his strategy if the other players maintain theirs. Most of multiplayer games tend to reach Nash equilibrium after a certain number of iterations.
	This area of study appeared with the aim of better understanding the economy, but it quickly extended to biology \cite{biologia}, politics \cite{politica}, computation and computer science \cite{computacion} (see review \cite{games}).

	In recent decades, physicists have begun to introduce features of the quantum world into game theory in order to gain advantages over classical strategies \cite{Meyers}-\cite{K18}. There are several reasons why quantifying games can be interesting. The first is simply because of the number of applications game theory has had in different fields. Moreover, its probabilistic nature makes one want to extend it to quantum probabilities. Another reason is the connection between game theory and quantum information theory. In fact, in the games themselves, players transmit information to other players and, since our world is quantum, it can be interpreted as quantum information \cite{Eisert}.
	
	There are several examples of well-known quantum game models. One is the prisoner's dilemma where, by exploiting the peculiarities of quantum behaviour, both players can escape from the dilemma \cite{Eisert}. Another case is the PQ game (flipping or not flipping a coin a certain number of times each player), where it has been shown that if the first player can use quantum strategies (superposition of both options) she/he will always win no matter what action the second player takes \cite{Meyers}. Not only have quantum games been proposed, but they are already being used to model, for example, human decision-making behaviour \cite{Y08}-\cite{Y20}.

	This paper deals with the well-known (in Spain) Chinos game, which traditionally consists of a group of people who hide a certain number of coins in their hands. The aim of each of them is to guess the total number of coins hidden by all of them. The Chinos  game is a variant 
	of the Morra game that is played with fingers instead of coins. It  dates back thousands of years to ancient Roman and Greek times \cite{morra}. 	
	This simple game shows a variety of behavioural patterns that have been used to model financial markets and information transmission \cite{Pastor}. 
	
	This is a non-cooperative game, in which each player will seek to maximize their chances of victory and minimize those of the other players. For this reason, throughout the work we will always look for this situation in the analysis of the possible strategies. In other words, the Nash equilibrium of the game model will be pursued. We must stress the importance of entanglement in quantum games when looking for the Nash equilibrium \cite{entrelazamiento, entrelazamiento2, entrelazamiento3}. It is worth mentioning that some  quantum games like the prisioner's dilemma and the PQ game have been already implemented using IBM quantum computers \cite{IBM, IBM2}.

	The organization of the paper is as follows. In section II we define in an abstract manner the Chinos game. 
	 In section III we review a classical Chinos game. In section IV we review the semiclassical model of the Chinos game proposed in \cite{GMD} and we propose two new semiclassical models. In section V we define a one-qubit game using unitary transformations. In section  VI
	 we introduce a two-qubit game that incorporates entanglement.  
	  In section VII we implement the latter game  using an IBM quantum computer, and finally, 
in section VIII	we state our conclusions and prospects.

\section{Abstract Chinos game for two players}

\begin{figure}[h!]
    \centering
    \includegraphics[width=8.5cm]{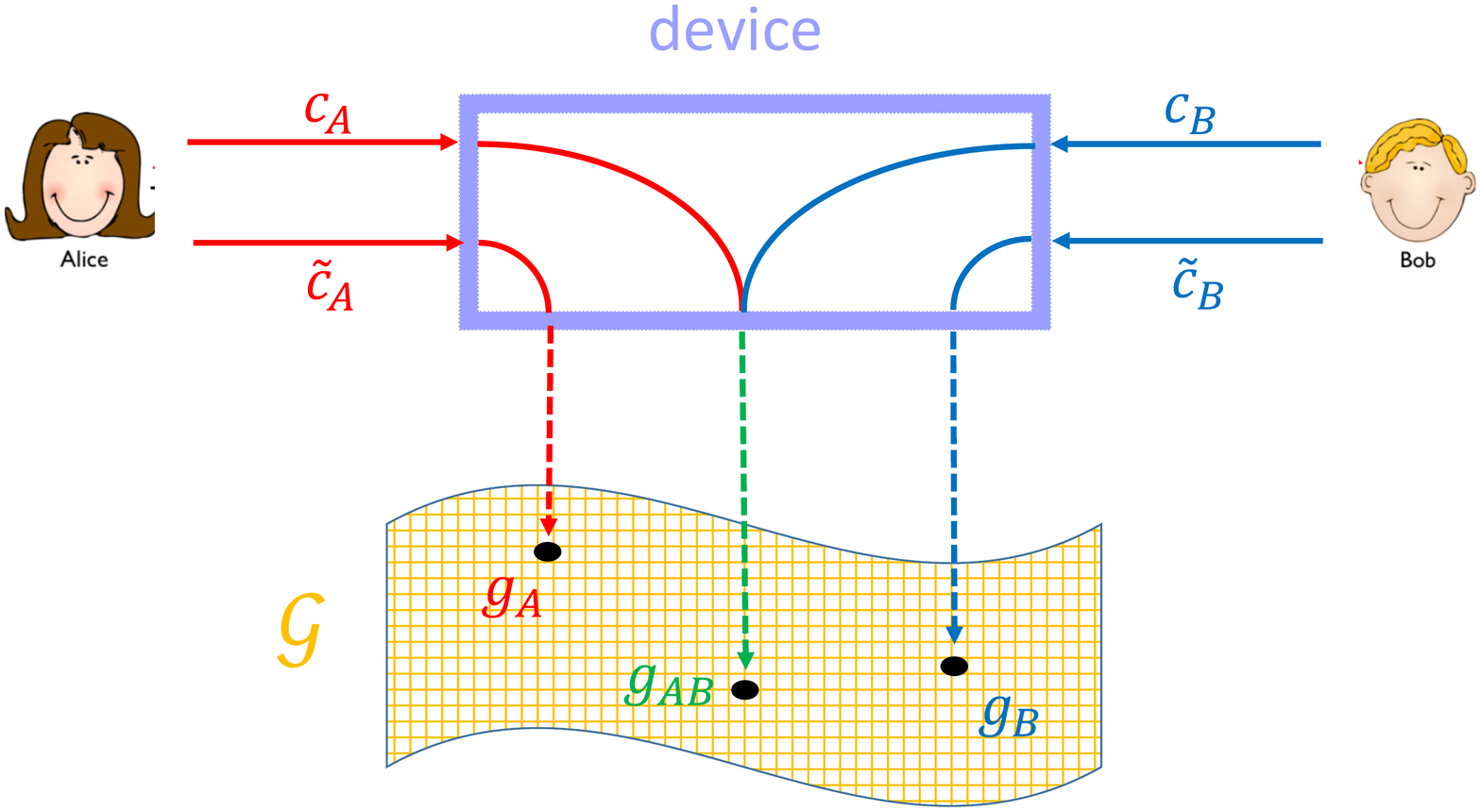}
 \caption{Protocol of the Chinos game}
  \label{fig-CHINO}
\end{figure}

Let Alice and Bob have access to a set of objects ${\cal C}$. 
At the start of the game, Alice chooses one object from ${\cal C}$, denoted $c_A$,
and Bob  chooses an object $c_B \in {\cal C}$. Both players are unaware of the other player's choice.
Alice and Bob then send $c_A$  and $c_B$  to a device that assigns them an object $g_{AB}$  of a set ${\cal G}$.
The goal of the players is to guess $g_{AB}$. To do so there is a  set ${\cal \tilde{C}}$ 
where  Alice and Bob selects the objects $\tilde{c}_A $ and $\tilde{c}_B$ respectively,  and send them to the previous device
that assigns them the objects $g_A$ and $g_B$ in ${\cal G}$. The protocol is shown in Fig. \ref{fig-CHINO}.

In order to quantify how good the players' guesses are, we shall use a distance $d(g, g')$  between objects $g, g' \in {\cal G}$,
that is, ${\cal G}$ is a metric space. 
The player whose guess has the shortest distance to the joint object will win. 
This means that both players will try to minimize the distances $d(g_A, g_{AB})$ and $d(g_B, g_{AB})$. 
In the Chinos game, Bob's guess must be different from Alice's. We shall call this the restriction rule. 
This is implemented by imposing a minimum distance between both guesses, that is $d(g_A, g_B) \geq d_{0} > 0$. 
To check this condition, Alice sends her guess $g_A$ to Bob, who calculates the distance with his guess $g_B$. 
Finally,  we shall suppose that Alice and Bob are intelligent agents. 
This implies that Alice's choice of $c_A$ can determine a set of choices of $\tilde{c}_A$ 
whose associated guesses $g_A$ may have some chance of winning. The same criteria apply to Bob.
The intelligence of the players is done by a map ${\cal I}: {\cal C} \rightarrow 2^{\cal \tilde{G}}$,
where  $2^{\cal \tilde{G}}$ is the set of all partitions of ${\cal \tilde{G}}$.

In summary,  the ingredients of the game are

\begin{itemize}

\item Player's choices: ${\cal C}$ and ${\cal \tilde{C}}$. 

\item  Player's guesses: ${\cal G}$.

\item Device:  choices $\rightarrow$ guesses.

\item Payoffs: distances.

\item Restriction rule: minimal distance.

\item Intelligence rule. 

%

\end{itemize}

A  consequence of the triangle inequality of the metric and the restriction placed of Bob's guesses is
\beq
d_0 \leq d(g_A, g_{B})  \leq d(g_A, g_{AB}) + d(g_B, g_{AB})  \, . 
\label{tri}
\eeq
This inequality implies that if the distance of Alice's conjecture
to  $g_{AB}$ is less than $d_0/2$ then  Bob's guess will be more than $d_0/2$ and therefore he will lose the game. 

We are going to present below several concrete realizations of the Chinos game. 
These can be classified into classical games or quantum games according to the 
classical or quantum nature of the space ${\cal G}$.

\section{Classical Chinos game}

	The game is defined by two parameters: $N_p$ that is the number of players and $N_c$ that is the number of coins that each player holds in her/his hand. In each round, each player draws from 0 to $N_c$ coins and guesses the total number of coins that they all draw, with the restriction  that the result predicted by the previous players cannot be repeated \cite{GMD}. 
	
We shall consider below the game with only  two players, $N_p = 2$, and one coin, $N_c = 1$. 
The best strategy for Alice is to choose randomly $c_A$ coins and to guess always $g_{A} =1$, so as not to reveal information to Bob \cite{GMD}. 
This is also based on the fact that with four  possible tosses the most likely value of the sum  is 1. 
Bob's  best strategy is to choose  $c_B$  at random and make his attempt $g_{B}$ in an ``intelligent'' way.
This  means that, if Bob chooses $c_B$=0, then he must exclude the option $g_{B}$=2 and if  he chooses $c_B$=1, 
he must exclude $g_B$=0. Table \ref{tabla1} shows all the possible options assuming that each player follows her/his best strategy. 

 \begin{table}[h!]
    \centering
    \begin{tabular}{|c|c|c|c|c|c|}
       \hline
      $c_A$ & $c_B$ & $g_{AB}$ & $g_A$ & $g_{B}$ & Winner\\
      \hline
        0 & 0 & 0 & 1 & 0 & B\\ 
   \hline
        0 & 1 & 1 & 1 & 2 & A\\ 
           \hline
        1 & 0 & 1 & 1 & 0 & A\\ 
       \hline
        1 & 1 & 2 & 1 & 2 & B\\ 
        \hline
    \end{tabular}
    \caption{\small{All possible options for the Chinos game with $N_p=2, N_c=1$.}}
    \label{tabla1}
\end{table}	
It is clear that  each player will win half of the time and therefore their winning probabilities are equal, that is
 \begin{equation}
    P_{A} = P_{B} = \frac{1}{2}.
    \label{AB}
\end{equation}
  Moreover, their strategies are stable. 
  In reference \cite{GMD} it was shown that \eqref{AB} also holds for two players and a generic number of coins $N_c$. 
 This was called  a classical symmetry between the players. These authors 
also proposed in \cite{GMD} a semiclassical and   quantum version of the Chinos game to test whether the classical symmetry \eqref{AB}  
is broken by quantum fluctuations. 
 
From the formal viewpoint introduced in section II,  this game involves the following sets: 
\beq
{\cal C} = \{ 0, 1 \}, \quad 	{\cal \tilde{C}} =  {\cal G} =  \{ 0, 1, 2 \} \, . 
  \label{CG}
\eeq	
where  ${\cal C}$ contains the number of coins, $c_A$ and $c_B$, of Alice and Bob, 
 ${\cal G}$ contains their guesses $g_A$ and $g_B$ for the total number of coins $g_{AB}$. 
 The device operates as follows
\beq
g_A = \tilde{c}_A , \quad 	g_B = \tilde{c}_B, \quad g_{AB} = c_A + c_B \, .
\label{dev}
\eeq
The guesses $g_A, g_B$ are identical to the choices $\tilde{c}_A, \tilde{c}_B$, 
but not in the games we shall consider below. 
The metric in ${\cal G}$ is given by 
\beq
d(g, g') = |g - g'|, \qquad g, g' \in {\cal G} \, . 
\label{dg}
\eeq
and the minimal distance is set to $d_0 =1$. The restriction rule amounts to
\barray 
{\rm if} \; g_A = 0 & \rightarrow g_B = 1,2  \; ,  \label{r1}  \\
{\rm if} \; g_A = 1 & \rightarrow g_B = 0,2 \; , \nonumber   \\
{\rm if} \; g_A = 2 & \rightarrow g_B = 0,1 \nonumber  \; ,    \nonumber 
\earray 
and  the intelligent rule to
\barray 
{\rm if} \; c_{A/B} = 0 & \rightarrow \tilde{c}_{A/B} = g_{A/B} =  0,1  \; ,  \label{i1}  \\
{\rm if} \; c_{A/B} = 1 & \rightarrow \tilde{c}_{A/B} = g_{A/B} =  1,2  \; .  \nonumber 
\earray

\section{Semiclassical models}

A semi-classical version of the Chinos  game consists of replacing coins with a quantum degree of freedom and  measuring an observable on it.
The value taken by  that observable is the goal of the players' guesses \cite{GMD}. We shall first review the proposal of reference \cite{GMD} 
that uses a single boson degree of freedom and later on  we shall propose two related  games.  


Let $b$ and $b^\dagger$ be the bosonic creation and  annihilation operators satisfying the canonical commutation rule [$b,b^{\dagger}$] = 1.
They  generate the  bosonic states in the standard way: $b|0\rangle$ = 0, $b^{\dagger}|n\rangle = \sqrt{n+1}|n+1\rangle$ where $|n\rangle = (b^{\dagger})^n |0\rangle / \sqrt{n! }$.

We will study the case of two players. Each of them can act on the bosonic state  $|0 \rangle$ with one of the following 
operators
\begin{equation}
    O_{1}=I,  O_{2}=\frac{I+b^{\dagger}}{\sqrt{2}}, 
  O_{3}=\frac{I-b^{\dagger}}{\sqrt{2}},  O_{4}=b^{\dagger}.
\label{eq8}
\end{equation}

This is the quantum analogue of drawing a coin. Let us denote by $O_i^A$ the operator chosen by Alice
and by $O_j^B$ the operator chosen by Bob. The joint state produced by both players is given by  


%
\begin{equation}
    |\Psi_{i,j}\rangle=N^{-1/2}_{i,j}O_{i}^{A}O_{j}^{B}|0\rangle=\sum_{n=0}^{2}c_{i,j}(n)|n\rangle,
    \label{eq6}
\end{equation}

\noindent where $N_{i,j}$ is a normalization constant. The operators 
$O_1$ and $O_4$, are equivalent to the classical choices of 0 and 1 coins respectively.
In this case  the state \eqref{eq6} is simply  $|0\rangle, |1\rangle $ or $|2 \rangle$, that
brings us back to  the classical game.  
Allowing Alice and Bob to use also  $O_2$ and $O_3$ generates linear superpositions
in \eqref{eq6}, that leads to probabilistic outcomes. They  will be  analyzed in terms 
of the  probability of finding the state $|n \rangle$ in \eqref{eq6} 
\begin{equation}
    p_{i,j}(n)=|\langle n|\Psi_{i,j}\rangle|^{2} = c_{i,j}^{2}(n).
   \label{cla10} 
\end{equation}

Suppose the both players follow the classical strategy, i.e. choose randomly among the operators $O_i$, $i=1,2,3,4$.
The probability of Alice guessing $n$ under the choice $O_i^A$ is given by the average over Bob's choices,

\begin{equation}
    \langle p_i(n) \rangle  = \frac{1}{4} \sum_{j=1}^4 p_{i,j}(n).
\end{equation}

The results are given in Table \ref{tabla3}.

\begin{table}[h!]
\centering
\begin{tabular}{|c|c|c|c|c|}
\hline
 & $O_{1}^{A}$ & $O_{2}^{A}$ & $O_{3}^{A}$ & $O_{4}^{A}$\\ \hline
  $\langle p(0) \rangle$  &  1/2 & 41/168  &   41/168 & 0    \\ \hline
 $\langle p(1) \rangle$ &  1/2 & 59/168  &   59/168 & 5/12 \\ \hline
 $\langle p(2) \rangle$  &  0 & 68/168  &   68/168 & 7/12 \\ \hline
 
\end{tabular}
 \caption{\small{Average odds of getting 0, 1 or 2 for Alice.}}
    \label{tabla3}
\end{table}

Hence if Alice draws $O_1^A$ her best  guess is 0 (or 1) and on drawing $O_2^A$, $O_3^A$ and $O_4^A$ her best guess is 2.
Therefore, the total probability of winning for Alice is 

\begin{equation}
    P_{A}= \frac{1}{4} \cdot \frac{1}{2} + \frac{1}{2} \cdot \frac{68}{168} + \frac{1}{4} \cdot \frac{7}{12} = \frac{53}{112} < \frac{1}{2}.
\end{equation}

After many  rounds Alice realizes that she is loosing the game and decides to change her strategy by randomly choosing between 
the operators $O_1^A$ and $O_4^A$. Her  probability of winning changes to 

\begin{equation}
    P_{A} = \frac{1}{2} \cdot \frac{1}{2} +\frac{1}{2} \cdot \frac{7}{12} = \frac{13}{24} >\frac{1}{2}.
\end{equation}

Bob then notices this imbalance and decides to choose randomly between the operators $O_1^B$ and $O_4^B$,
which reproduces the classical game where the classical symmetry \eqref{AB} is restored \cite{GMD}.  

The formalization of the game is as follows.
Alice and Bob choose $i, j \in {\cal C}$ where 
\beq 
{\cal C} = \{ 1,2,3,4 \}, 
\label{b14}
\eeq
 and send them to the device that constructs the state \eqref{eq6}
belonging to the Hilbert space expanded by the boson states with 0, 1 and 2 occupancies, 
\beq
{\cal G} = {\rm Span} \{ |0 \rangle, |1 \rangle, |2 \rangle \} \ . 
\label{b15}
\eeq
Each  player's  guess of  these occupancies, $n$, is mapped by  device into the basis of ${\cal G}$  as 
\beq
n \in {\cal \tilde{C}} =  \{ 0, 1, 2 \}  \rightarrow  |n \rangle \in {\cal G}  \, . 
\label{b16} 
\eeq
To define a distance between two  states  $| g \rangle, |g' \rangle \in {\cal G}$, 
we use the trace distance between the corresponding density matrices, 
$\rho_g = | g \rangle \langle g|$ and $\rho_{g'} =  | g' \rangle \langle g'|$  \cite{nielsen} 
\beq
d( | g \rangle, |g' \rangle ) =  \frac{1}{2} {\rm tr} | \rho_g - \rho_{g'} | = \sqrt{ 1- | \langle g | g' \rangle|^2}   \, . 
\label{b17}
\eeq

The probability \eqref{cla10} is related to this distance as
\beq
d( |n \rangle, |\Psi_{i,j} \rangle) = \sqrt{ 1 -  p_{i,j} (n)  } .
\label{b18} 
\eeq

Hence minimizing the distance of the guess $| n \rangle$ to the state  $|\Psi_{i,j} \rangle$
is equivalent  to maximizing the probability $p_{i,j} (n)$. The latter quantity is equal to 
the square of the fidelity between these states.

\subsection{Hard-core boson}

We shall next propose a model where the boson operator $b^\dagger$, used previously,  is replaced by a hard-core boson,
that is, an operator satisfying the condition:
\begin{equation}
    (b^{\dagger})^{2} = 0.
\end{equation}

This forces the  elimination of  the operator $O_4$ as an option since its use by Alice and Bob would  lead to a null move. 
This also  implies that the classical version of the chinos game is not included in the new version.

The study of strategies is based again on the probabilities of obtaining 0 and 1 for each possible move. They  are given in Table \ref{tabla4}.

\begin{table}[h!]
\centering
\begin{tabular}{|c|c|c|c|}
\hline
  & $O_{1}^{A}$ & $O_{2}^{A}$ & $O_{3}^{A}$ \\ \hline
 $O_{1}^{B}$ & \begin{tabular}[c]{@{}c@{}}$p(0)$ = 1\\ $p(1)$ = 0\end{tabular}    & \begin{tabular}[c]{@{}c@{}}$p(0)$ = 1/2\\ $p(1)$ = 1/2\end{tabular}   & \begin{tabular}[c]{@{}c@{}}$p(0)$ = 1/2\\ $p(1)$ =1/2\end{tabular}     \\ \hline
 $O_{2}^{B}$ & \begin{tabular}[c]{@{}c@{}}$p(0)$ =1/2\\ $p(1)$ = 1/2\end{tabular} & \begin{tabular}[c]{@{}c@{}}$p(0)$ = 1/5\\ $p(1)$ = 4/5\end{tabular} & \begin{tabular}[c]{@{}c@{}}$p(0)$ = 1\\ $p(1)$ = 0\end{tabular}   \\ \hline
 $O_{3}^{B}$ & \begin{tabular}[c]{@{}c@{}}$p(0)$ =1/2\\ $p(1)$ = 1/2\end{tabular} & \begin{tabular}[c]{@{}c@{}}$p(0)$ = 1\\ $p(1)$ = 0\end{tabular}   & \begin{tabular}[c]{@{}c@{}}$p(0)$ = 1/5\\$ p(1)$ = 4/5\end{tabular} \\ \hline
\end{tabular}
 \caption{\small{Odds of getting 0 or 1 for all possible moves in the game with hard-core bosons.}}
    \label{tabla4}
\end{table}

Following  the classical strategy, the players will draw at random among the operators $O_i$, $i=1,2,3$. 
Averaging  over Bob's choices,  the probabilities of Alice getting 0 and 1 are given in Table \ref{tabla5}.

\begin{table}[h!]
\centering
\begin{tabular}{|c|c|c|c|}
\hline
 & $O_{1}^{A}$ & $O_{2}^{A}$ & $O_{3}^{A}$ \\ \hline
  $\langle p(0) \rangle$  &  2/3 & 17/30 &   17/30    \\ \hline
 $\langle p(1) \rangle$ &  1/3 & 13/30  &   13/30 \\ \hline
 \end{tabular}
 \caption{\small{Average odds of getting 0, 1 or 2 for Alice.}}
    \label{tabla5}
\end{table}

It is clear that Alice's best guess is 0,  regardless her choice of operator, with a winning probability

\begin{equation}
    P_{A} = \frac{1}{3} \cdot \frac{2}{3} +\frac{2}{3} \cdot \frac{17}{30} = \frac{3}{5} >\frac{1}{2}.
\end{equation}

To reverse this outcome Bob will choose randomly between the $O_2$ and $O_3$ that seem more favorable
from Table \ref{tabla5}. However, Alice's winning probability is still higher than a half, 
\begin{equation}
     \frac{1}{2} < P_{A} = \frac{17}{30} < \frac{3}{5}.
\end{equation}

Hence, Alice has achieved a winning and stable strategy that breaks the symmetry of the classical game.

To improve the chances of Bob we shall use the following operators

\begin{equation}
    O_{1} = I, O_{2} = cI + sb^{\dagger}, O_{3} = cI - sb^{\dagger},
    \label{eq15}
\end{equation}

\noindent where $c = \cos(\theta), s = \sin(\theta)$ with $\theta \in (0,\pi/2)$. The previous case corresponds to  $\theta = \pi/4$. 
The values $\theta = 0, \pi/2$ are excluded because in these cases $O_2 = \pm O_3$.  

The probabilities of obtaining 0 or 1 in the boson occupation are shown in Table \ref{tabla6}.

\begin{table}[h!]
\centering
\begin{tabular}{|c|c|c|c|}
\hline
  & $O_{1}^{A}$ & $O_{2}^{A}$ & $O_{3}^{A}$ \\ \hline
  $O_{1}^{B}$ & \begin{tabular}[c]{@{}c@{}}$p(0)$ = 1\\ $p(1)$ = 0\end{tabular}    & \begin{tabular}[c]{@{}c@{}}$p(0)$ = $c^{2}$\\ $p(1)$ = $s^{2}$\end{tabular}   & \begin{tabular}[c]{@{}c@{}}$p(0)$ = $c^{2}$\\ $p(1)$ =$s^{2}$\end{tabular}     \\ \hline
 $O_{2}^{B}$ & \begin{tabular}[c]{@{}c@{}}$p(0)$ =$c^{2}$\\ $p(1)$ = $s^{2}$\end{tabular} & \begin{tabular}[c]{@{}c@{}}$p(0)$ = $\frac{c^{2}}{c^{2} + 4s^{2}}$\\ $p(1)$ = $\frac{4s^{2}}{c^{2} + 4s^{2}}$\end{tabular} & \begin{tabular}[c]{@{}c@{}}$p(0)$ = 1\\ $p(1)$ = 0\end{tabular}   \\ \hline
 $O_{3}^{B}$ & \begin{tabular}[c]{@{}c@{}}$p(0)$ =$c^{2}$\\ $p(1)$ = $s^{2}$\end{tabular} & \begin{tabular}[c]{@{}c@{}}$p(0)$ = 1\\ $p(1)$ = 0\end{tabular}   & \begin{tabular}[c]{@{}c@{}}$p(0)$ = $\frac{c^{2}}{c^{2} + 4s^{2}}$\\ $p(1)$ = $\frac{4s^{2}}{c^{2} + 4s^{2}}$\end{tabular} \\ \hline
\end{tabular}
 \caption{Odds of getting 0 or 1 for all possible moves as a function of $\theta$.}
    \label{tabla6}
\end{table}

\begin{figure}[h!]
    \centering
    \includegraphics[width=8cm, height=6cm]{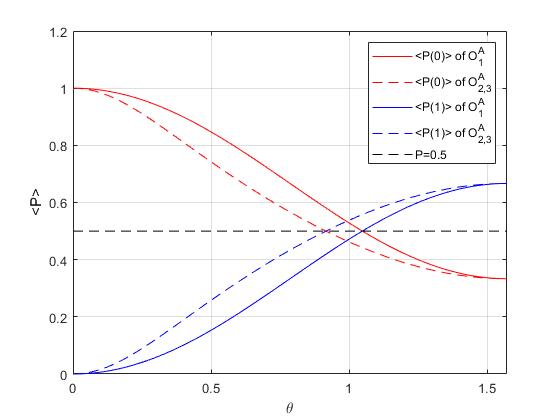}
    \caption{Plot of the probabilities of Table \ref{tabla7}}
    \label{figure1}
\end{figure}

Applying the classical strategies, the probabilities for Alice averaged over Bob's choices are given in Table \ref{tabla7}
and plotted in Fig. \ref{figure1}. 

\begin{table}[h!]
\centering
\begin{tabular}{|c|c|c|}
\hline
 & $O_{1}^{A}$ & $O_{2,3}^{A}$ \\ \hline
  $\langle p(0) \rangle$   &  $\frac{1}{3}(1+2c^2)$ & $\frac{2c^{2} + 4s^{2} + c^{4} + 4c^{2}s^{2}}{3(c^{2} + 4s^{2})}$  \\ \hline
 $\langle p(1) \rangle$  &  $\frac{2}{3} s^2$ & $\frac{s^{2}c^{2} + 4s^{4} + 4s^{2}}{3(c^{2} + 4s^{2})}$  \\ \hline
 \end{tabular}
 \caption{\small{ Average odds of getting 0 or 1 for Alice.}}
    \label{tabla7}
\end{table}

For every Alice's strategy, the  probabilities $\langle p(n) \rangle (n=0,1)$  
cross at $1/2$ at the angles $\theta_1 < \theta_2$ given by  (the probability of  $O_2^A$ and $O_3^A$ are the same)
\barray 
 \langle p(0) \rangle_{O_2^A} & =  &  \langle p(1) \rangle_{O_2^A} = \frac{1}{2} \Longrightarrow \theta_{1} = 0.9155.. \label{theta} \\
        \langle p(0) \rangle_{O_1^A}   & =  &     \langle p(1) \rangle_{O_1^A} = \frac{1}{2}    \Longrightarrow  \theta_2 = 1.0472.. \nonumber 
\earray

Consequently the strategies followed by Alice depend on the value of $\theta$ as follows

\begin{itemize}
    \item 0 $< \theta < \theta_{1}$:  Alice will guess 0 for all  $O_{i}^{A}$.
    \item $\theta_{1} < \theta < \theta_{2}$ :  Alice will guess 0 for  $O_{1}^{A}$, and 1 for $O_{2}^{A}$ or $O_{3}^{A}$.
    \item $\theta_{2} < \theta < \pi/2$ :  Alice will guess 1 for all  $O_{i}^{A}$.
    
\end{itemize}

Averaging the corresponding probabilities yields a value of $P_A$ larger than $1/2$  for  all $\theta's$,
so reproducing the previous case where $\theta = \pi/4$. An analysis of the strategies that Bob may adopt
does not change the situation even at the values of $\theta_1$ and $\theta_2$.   
We conclude that Alice always has a winning and stable strategy in this game.
 
The sets ${\cal C}, {\cal \tilde{C}}$ and the space $ {\cal G}$ involved in the hard-core boson game are simply the truncation of  those used in the boson game, namely 
\barray 
{\cal C} & =  &  \{ 1,2,3 \}, \;  {\cal \tilde{C}} =  \{ 0, 1 \} ,\;  {\cal G} = {\rm Span} \{ |0 \rangle, |1 \rangle \},  
\label{b24}
\earray 
while the distance for ${\cal G}$ is the same as  in\eqref{b18}.

\section{One-qubit  games}

In previous games, the players used the operators $O_i$ that are not unitary, which makes 
it difficult to carry out experimentally. In this section and the next one  we are going to propose two games that solve this difficulty.
In the first, game we shall replace the hard-core boson by a qubit initialized in the state 
$|0 \rangle$ on which the players act choosing between the following unitary transformations
\begin{equation}
O_1= I, \quad O_2 = e^{ - \frac{i \theta}{2} \sigma^y} ,\quad O_3 = e^{  \frac{i \theta}{2} \sigma^y}  \, , 
\label{1q1}
\end{equation}
with  $\theta \in (0, \pi/2)$. The operators $O_2$ and $O_3$ are rotations of angles $\theta$ and $-\theta$ around the $y$-axis. 
The state constructed by the device with the information provided by Alice and Bob is
\beq
|\Psi_{i,j} \rangle =  O^A_i O^B_j | 0 \rangle
\label{1q2}
\eeq
This state is already normalized since the operators \eqref{1q1} are unitary.

The probabilities of obtaining the states $|0\rangle$ or $|1\rangle$ are shown in Table \ref{tabla9}.
Notice the similarities with  Table \ref{tabla6}.

\begin{table}[h!]
\centering
\begin{tabular}{|c|c|c|c|}
\hline
  & $O_{1}^{A}$ & $O_{2}^{A}$ & $O_{3}^{A}$ \\ \hline
  $O_{1}^{B}$ & \begin{tabular}[c]{@{}c@{}}$p(0)$ = 1\\ $p(1)$ = 0\end{tabular}    & \begin{tabular}[c]{@{}c@{}}$p(0)$ = $c^2$\\ $p(1)$ = $s^2$\end{tabular}   & \begin{tabular}[c]{@{}c@{}}$p(0)$ = $c^2$\\ $p(1)$ =$s^2$\end{tabular}     \\ \hline
 $O_{2}^{B}$ & \begin{tabular}[c]{@{}c@{}}$p(0)$ =$c^2$\\ $p(1)$ = $s^2$\end{tabular} & \begin{tabular}[c]{@{}c@{}}$p(0)$ = $c^2$\\ $p(1)$ = $s^2$\end{tabular} & \begin{tabular}[c]{@{}c@{}}$p(0)$ = 1\\ $p(1)$ = 0\end{tabular}   \\ \hline
 $O_{3}^{B}$ & \begin{tabular}[c]{@{}c@{}}$p(0)$ =$c^2$\\ $p(1)$ = $s^2$\end{tabular} & \begin{tabular}[c]{@{}c@{}}$p(0)$ = 1\\ $p(1)$ = 0\end{tabular}   & \begin{tabular}[c]{@{}c@{}}$p(0)$ = $c^2$\\ $p(1)$ = $s^2$\end{tabular} \\ \hline
\end{tabular}
 \caption{
 $c= \cos(\theta/2)$ and $s = \sin(\theta/2)$. }
    \label{tabla9}
\end{table}

\begin{figure}[h!]
    \centering
    \includegraphics[width=8cm, height=6cm]{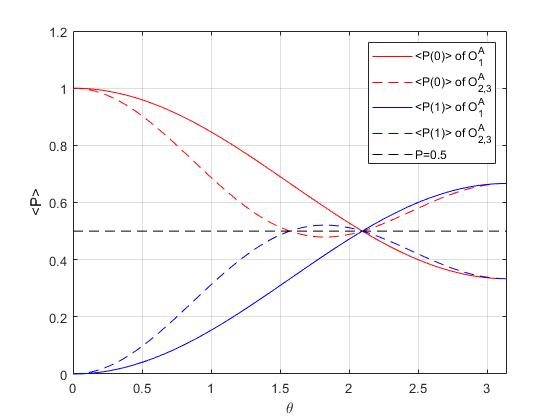}
    \caption{Plot of probabilities given in Table \ref{tabla10}.}
    \label{figure2}
\end{figure}

Applying the classical strategies, the probabilities for Alice averaged over Bob's choices are given in Table \ref{tabla10}
and plotted in Fig. \ref{figure2}.

\begin{table}[h!]
\centering
\begin{tabular}{|c|c|c|}
\hline
 & $O_{1}^{A}$ & $O_{2,3}^{A}$ \\ \hline
  $\langle p(0) \rangle$   &  $\frac{1}{3}(1+2 c^2)$ & $\frac{1}{3}(2 - 3 c^2 + 4 c^4)$    \\ \hline
 $\langle p(1) \rangle$  &  $\frac{2}{3}s^2$ & $\frac{1}{3}(1 + 3 c^2 - 4 c^4)$  \\ \hline
 \end{tabular}
 \caption{\small{ Average odds for Alice.}}
   \label{tabla10}
\end{table}

As in the previous game there are three regions where Alice's strategy is winning and stable. 
However, at their boundaries, Bob is able to balance the game, restoring  the classical symmetry. 
The values of $\theta$ at the boundaries are given by 

\barray 
    \langle p(0) \rangle_{O_{2,3}^A} & =  &   \langle p(1) \rangle_{O_{2,3}^A} = \frac{1}{2}  \Longrightarrow \theta_{1}' = \frac{\pi}{2},  \label{thetap}\\
  \langle p(0) \rangle_{O_{1,2,3}^A} & =  &  \langle p(1) \rangle_{O_{1,2,3}^A} = \frac{1}{2} \Longrightarrow \theta_{2}' = \frac{2\pi}{3}.
\nonumber
\earray

The strategies followed by Alice are the following:

\begin{itemize}
     \item 0 $\leq \theta < \theta_{1}'$ : Alice will guess 0 regardless of her choice of operator between $O_{1}^{A}$, $O_{2}^{A}$ and $O_{3}^{A}$.
    \item $\theta_{1}' < \theta < \theta_{2}'$ :Alice will guess 0 if she chooses $O_{1}^{A}$ and 1 if she chooses $O_{2}^{A}$ or $O_{3}^{A}$.
    \item $\theta_{2}' < \theta < \pi/2$ : Alice will guess 1 if she chooses $O_{1}^{A}$ and 0 if she chooses $O_{2}^{A}$ or $O_{3}^{A}$.
\end{itemize}

At the boundary points we have the following situations: 

\begin{itemize}
    \item $\theta = \theta_{1}'$: Bob notices that  choosing randomly between $O_{2}^{B}$ and $O_{3}^{B}$ the game becomes symmetric 
    for all Alice's draws.  Table \ref{tabla11} shows the corresponding probabilities.  
    Alice has therefore a winning but unstable strategy as Bob can balance it.

    \begin{table}[h!]
\centering
\begin{tabular}{|c|c|c|c|}
\hline
  & $O_{1}^{A}$ & $O_{2}^{A}$ & $O_{3}^{A}$ \\ \hline 
 $O_{2}^{B}$ & \begin{tabular}[c]{@{}c@{}}$p(0)$ = 1/2\\ $p(1)$ = 1/2\end{tabular} & \begin{tabular}[c]{@{}c@{}}$p(0)$ = 0\\ $p(1)$ = 1\end{tabular} & \begin{tabular}[c]{@{}c@{}}$p(0)$ = 1\\ $p(1)$ = 0\end{tabular}   \\ \hline
 $O_{3}^{B}$ & \begin{tabular}[c]{@{}c@{}}$p(0)$ =1/2\\ $p(1)$ = 1/2\end{tabular} & \begin{tabular}[c]{@{}c@{}}$p(0)$ = 1\\ $p(1)$ = 0\end{tabular}   & \begin{tabular}[c]{@{}c@{}}$p(0)$ = 0\\ $p(1)$ = 1\end{tabular} \\ \hline
\end{tabular}
 \caption{\small{Values given in Table \ref{tabla9} for $\theta'_1$ and Bob's choices $O_{2}^{B}$ and $O_{3}^{B}$}}
    \label{tabla11}
\end{table}

    \item $\theta = \theta_{2}'$ : 
 At this point all the curves in Fig. \ref{figure2} intersect at the value of 1/2. 
 Bob does not even have to change his strategy to make the game symmetrical again. Moreover, Alice can do nothing to prevent this. This is observed by evaluating the probabilities shown in Table \ref{tabla9} at $\theta_{2}'$.

\end{itemize}

We conclude that in the one-qubit Chinos game with unitary operators, Alice will have a winning and stable strategy for any angle $\theta \in (0, \pi)$ except at the values $\theta_{1}'$ and $\theta_{2}'$. In the former, Bob can choose a strategy that balances the game, and in the latter, the game is symmetrized without the need for Bob to change his strategy. This result is essentially the same as the one obtained using a hard-core boson, where Alice has always a winning stable strategy. 

In this one-qubit game, the  sets ${\cal C}, {\cal \tilde{C}}$ and the space $ {\cal G}$ are
the same as those of the hard-core boson given in eq.\eqref{b24}.

\section{Two-qubits games }

In the previous quantum games the players' guesses  where mapped by the device
onto the states belonging to the orthonormal basis of the Hilbert space ${\cal G}$. Along the lines of reference \cite{GMD},  we  shall next propose 
games where this condition is not imposed.   
This is implemented as follows. Alice  and Bob will guess the states 
$|g^{A}\rangle$ and $|g^{B}\rangle$ respectively, with the condition that Bob's state  should be orthogonal to Alice's state, that is 
\begin{equation}
     \langle g^{A}|g^{B}\rangle = 0.
     \label{eq21}
\end{equation}

In view of eq.\eqref{b17}, this implies that the distance between the players's guess is exactly 1, i.e. $d_0 =1$. 
This condition is the quantum version of the classical rule where players cannot repeat the guesses of the previous ones.
 Alice and Bob payoffs evaluate how successful each player was in predicting the state 
 generated jointly $|\Psi_{a,b} \rangle$,

\begin{equation}
     f^{A} = |\langle g^{A}|\Psi_{a,b}\rangle|^{2},\quad 
      f^{B} = |\langle g^{B}|\Psi_{a,b}\rangle|^{2} \, . 
      \label{fAB}
\end{equation}

Eq.\eqref{b17} relates these quantities to the distance of the players' guess to the joint state. 
The winner of the game is the one with the highest payoff.

We shall consider a two qubit system on the computational basis 
$|i_{1},i_{0}\rangle$ with $i_{0},i_{1} = 0,1$. The operators that generates the joint
state $|\Psi_{a,b} \rangle$ will be those that produce the  Bell states acting on $|0,0\rangle$.
The Bell states form an orthonormal basis of maximally entangled states 
of a two-qubit system given by

\begin{equation}
    |\phi^{\pm}\rangle = \frac{|00\rangle \pm  |11\rangle}{\sqrt{2}},  \quad |\psi^{\pm}\rangle = \frac{|10\rangle \pm  |11\rangle}{\sqrt{2}} \, . 
\end{equation}
They  can be constructed as 
\begin{equation}
    \begin{split}
    |\phi^{+}\rangle = Bell(0,0)|00\rangle,  \quad |\phi^{-}\rangle = Bell(0,1)|00\rangle, \\
      |\psi^{+}\rangle = Bell(1,0)|00\rangle,  \quad |\psi^{-}\rangle = Bell(1,1)|00\rangle.
      \end{split}
\end{equation}
with 
\begin{equation}
    Bell(i_{1},i_{0}) = CNOT(id \otimes H)(X^{i_{1}} \otimes X^{i_{0}}) \, , 
    \label{eq28}
\end{equation}
and where $H$ is the Hadamard gate, $X$ the  NOT gate and $CNOT$ the gate

\begin{equation}
    CNOT |i_{1},i_{0}\rangle = |i_{1} \oplus i_{0}, i_{0}\rangle,
\end{equation}
where  $i_0$ is the control qubit and $i_1$ is the target qubit.

We shall use in what follows the notation

\begin{equation}
\begin{split}
      O_{0} = Bell(0,0),  \quad & O_{1} = Bell(0,1),  \\
      O_{2} = Bell(1,0),  \quad & O_{3} = Bell(1,1).
    \end{split}
    \label{eq30}
\end{equation}

Using these operators Alice and Bob will  generate the joint state 
\begin{equation}
    |\psi_{a_{0},b_{0}}\rangle = O^{A}_{a_{0}}O^{B}_{b_{0}}|00\rangle \, , 
    \label{eq31}
\end{equation}
that is normalized because the operators \eqref{eq30} are unitary.

An interesting property of all the joint states \eqref{eq31} is that they are maximally entangled. 
This can be proved computing the reduced density matrix of the qubit 0 

\begin{equation}
    \rho_1 = \frac{1}{2}\begin{pmatrix}
    1 & 0\\
    0 & 1
    \end{pmatrix} \, . 
    \quad
    \label{rho} 
\end{equation}

Alice's and Bob's guesses are chosen as 

\barray 
  |g^{A}_{a_{1},a_{2}}\rangle &  =  &  O^{A}_{a_{1}}O^{A}_{a_{2}}|00\rangle; \quad a_{1},a_{2}=0,1,2,3 \, ,   \label{eq39} \\
   |g^{B}_{b_{1},b_{2}}\rangle & =  &  O^{B}_{b_{1}}O^{B}_{b_{2}}|00\rangle; \quad b_{1},b_{2}=0,1,2,3 \,  , \nonumber 
   \earray 
that  should be orthogonal to satisfy \eqref{eq21},   
\begin{equation}
     \langle g^{A}_{a_{1},a_{2}}|g^{B}_{b_{1},b_{2}}\rangle = 0.
     \label{eq41}
\end{equation}

The payoffs of the players are given by eq.\eqref{fAB} with $a,b$ replaced by $a_0, b_0$. 

It is convenient to define a metric in the space expanded by the operators $O_i O_j$ acting on $|00\rangle$ \cite{GMD},
\begin{equation}
\begin{split}
    G_{(i_{1},j_{1}),(i_{2},j_{2})} & = \langle 00|O^{\dagger}_{j_{1}}O^{\dagger}_{i_{1}}O_{i_{2}}O_{j_{2}}|00\rangle \, , 
    \end{split}
    \label{eq43}
\end{equation}
where $i_{1},j_{1},i_{2},j_{2} = 0,1,2,3$. In terms of this metric  the condition eq. (\ref{eq41}) becomes

\begin{equation}
     G_{(a_{1},a_{2}),(b_{1},b_{2})} = 0,
     \label{eq44}
\end{equation}
and the payoffs of each player,

\begin{equation}
\begin{split}
     f^A_{(a_1, a_2),(a_0, b_0)}= |G_{(a_{1},a_{2}),(a_{0},b_{0})}|^{2},\\
     f^B_{(b_1, b_2),(a_0, b_0)} = |G_{(b_{1},b_{2}),(a_{0},b_{0})}|^{2}.
\end{split}
\label{eq45}
\end{equation}

\begin{table}[h!]
\centering
\resizebox{8cm}{!}{
\begin{tabular}{|r|r|r|r|r|r|r|r|r|r|r|r|r|r|r|r|r|}
\hline
 $G$    & 00 & 22 & 12 & 30 & 13 & 31 & 01 & 23 & 02 & 20 & 10 & 32 & 11 & 33 & 03 & 21 \\ \hline
00 & 1  & 1  & $1$  & $1$  & $0$  & $0$  & 0  & 0  & 0  & 0  & 0  & 0  & 0  & 0  & 0  & 0  \\ \hline
22 & 1  & 1  & $1$  & $1$  & $0$  & $0$  & 0  & 0  & 0  & 0  & 0  & 0  & 0  & 0  & 0  & 0  \\ \hline
12 & $1$  & $1$  & 1  & 1  & 0  & 0  & $0$  & $0$  & 0  & 0  & 0  & 0  & 0  & 0  & 0  & 0  \\ \hline
30 & $1$  & $1$  & 1  & 1  & 0  & 0  & $0$  & $0$  & 0  & 0  & 0  & 0  & 0  & 0  & 0  & 0  \\ \hline
13 & $0$  & $0$  & 0  & 0  & 1  & 1  & $-1$  & $-1$  & 0  & 0  & 0  & 0  & 0  & 0  & 0  & 0  \\ \hline
31 & $0$  & $0$  & 0  & 0  & 1  & 1  & $-1$  & $-1$  & 0  & 0  & 0  & 0  & 0  & 0  & 0  & 0  \\ \hline
01 & 0  & 0  & $0$  & $0$  & $-1$  & $-1$  & 1  & 1  & 0  & 0  & 0  & 0  & 0  & 0  & 0  & 0  \\ \hline
23 & 0  & 0  & $0$  & $0$  & $-1$  & $-1$  & 1  & 1  & 0  & 0  & 0  & 0  & 0  & 0  & 0  & 0  \\ \hline
02 & 0  & 0  & 0  & 0  & 0  & 0  & 0  & 0  & 1  & 1  & $1$  & $1$  & $0$  & $0$  & 0  & 0  \\ \hline
20 & 0  & 0  & 0  & 0  & 0  & 0  & 0  & 0  & 1  & 1  & $1$  & $1$  & $0$  & $0$  & 0  & 0  \\ \hline
10 & 0  & 0  & 0  & 0  & 0  & 0  & 0  & 0  & $1$  & $1$  & 1  & 1  & 0  & 0  & $0$  & $0$  \\ \hline
32 & 0  & 0  & 0  & 0  & 0  & 0  & 0  & 0  & $1$  & $1$  & 1  & 1  & 0  & 0  & $0$  & $0$  \\ \hline
11 & 0  & 0  & 0  & 0  & 0  & 0  & 0  & 0  & $0$ & $0$  & 0  & 0  & 1  & 1  & $-1$  & $-1$  \\ \hline
33 & 0  & 0  & 0  & 0  & 0  & 0  & 0  & 0  & $0$ & $0$  & 0  & 0  & 1  & 1  & $-1$  & $-1$  \\ \hline
03 & 0  & 0  & 0  & 0  & 0  & 0  & 0  & 0  & 0  & 0  & $0$  & $0$  & $-1$  & $-1$  & 1  & 1  \\ \hline
21 & 0  & 0  & 0  & 0  & 0  & 0  & 0  & 0  & 0  & 0  & $0$  & $0$  & $-1$  & $-1$  & 1  & 1  \\ \hline
\end{tabular}}
\caption{\small{Metric $G_{(i_{1},j_{1}),(i_{2},j_{2})}$} defined in eq. (\ref{eq43}) using the operators of eq. (\ref{eq30}) organized in the  sets \eqref{eq46}.}
 \label{tabla13}
\end{table}

The entries of the metric \eqref{eq43} are given in Table \ref{tabla13} and has a block diagonal structure in terms of the   
following  subsets,

	    \begin{equation}
	        \begin{split}
	            Set_{1} = \{00,22,12,30\} ,  & \quad Set_{2} = \{13,31,01,23\} , \\
	        Set_{3} = \{02,20,10,32\}  , & \quad Set_{4} = \{11,33,03,21\}.
	        \end{split}
	        \label{eq46}
	    \end{equation}
 
 Since  these sets are mutually orthogonal, if Alice chooses a state belonging to one set, 
then  Bob has to choose a state belonging to a different  set to satisfy condition \eqref{eq44}.

With this metric,  $f^A$  and $f^B$ can be either 0 or 1, that is  Alice and Bob can either win or lose.
In particular,  for any of Alice's choices, she can win with a certain Bob's  choice of $b_0$, i.e.
\beq 
\forall a_0, a_1, a_2, \exists \,  b_0 \;  /  \; 
f^A_{(a_1, a_2),(a_0, b_0)} =   1 \ .
\label{fA}
\eeq

The situation of Bob is different. His choice of $b_0$ reduces  the possible winning guesses to the following sets
\barray
{\rm if}  \; b_0 & = &  0 \;  {\rm or} \;   2 \Longrightarrow (b_1, b_2) \in Set_1 \;  {\rm or} \; Set_3 \, , 
\label{bSet} \\
{\rm if}  \; b_0 & = &  1 \;  {\rm or} \;  3 \Longrightarrow (b_1, b_2) \in Set_2 \;  {\rm or} \; Set_4 
\nonumber 
\nonumber 
\earray 
as can be verified in Table \ref{tabla13}.

Let us   consider an example. 	
Suppose that   Alice chooses at random  among the four sets of eq.~ \eqref{eq46} and that Bob chooses $b_{0} = 0$.
We shall consider the  following cases:
	\begin{itemize}
	    \item Alice selects $(a_1, a_2) \in Set_{1}$ then Bob, applying eq.\eqref{bSet}  and the orthogonality rule \eqref{eq44}, 
	    will choose $(b_1, b_2) \in Set_{3}$.  The payoffs \eqref{eq45} for $a_0=0,1,2,3$ and $b_0=0$ are given by 

\begin{table}[h!]
\centering	    
\begin{tabular}{|r|r|r|r|r|}
\hline 
$f^A$ & 00 & 10 & 20 &  30 \\ \hline
$Set_1$ & 1 & 0 & 0 & 1 \\
\hline
\end{tabular} \quad
\begin{tabular}{|r|r|r|r|r|}
\hline 
$f^B$ & 00 & 10 & 20 &  30 \\ \hline
$Set_3$ & 0 & 1 & 1 & 0 \\
\hline
\end{tabular}
\end{table}
	    
	     \item Alice selects $(a_1, a_2) \in Set_{2}$ then Bob 
	    choosing $(b_1, b_2) \in Set_{1} 	\cup Set_{3}  $.  Their payoffs  are  
	    
	    \begin{table}[h!]
\centering	    
\begin{tabular}{|r|r|r|r|r|}
\hline 
$f^A$ & 00 & 10 & 20 &  30 \\ \hline
$Set_2$ & 0 & 0 & 0 & 0 \\
\hline
\end{tabular} \quad
\begin{tabular}{|r|r|r|r|r|}
\hline 
$f^B$ & 00 & 10 & 20 &  30 \\ \hline
$Set_1$ & 1 & 0 & 0 & 1 \\ \hline
$Set_3$ & 0 & 1 & 1 & 0 \\
\hline
\end{tabular}
\end{table}

\item Alice selects $(a_1, a_2) \in Set_{3}$. This case is similar  to the choice of $Set_{1}$.
\item Alice selects $(a_1, a_2) \in Set_{4}$. This case is similar to the choice of  $Set_{2}$.
\end{itemize}

 The  average payoffs of Alice and Bob are given by
 \barray
 \langle f^A \rangle & = & \frac{1}{2} \left(  \frac{ 2}{4} + 0   \right) = \frac{ 1}{4} \, , 
 \label{ave1} \\
  \langle f^B \rangle & = & \frac{1}{2} \left(  \frac{ 2}{4} + \frac{ 2}{4} \right) = \frac{ 1}{2} \, , 
  \nonumber 
 \earray 
 that gives  the normalized probabilities 

	\begin{equation}
	    P_{A} = \frac{1}{3},  \quad P_{B} = \frac{2}{3}. 
	    \label{Ptot} 
	\end{equation}

	In view of this result, Alice will try to reverse the situation, but it will not be possible since, according to eq. \eqref{fA}, 
	given  $a_{0}$, any set can make her win depending on the value of $b_0$.
	We conclude that in this Chinos game Bob has a winning and stable strategy. This is in contrast with 
with game using  bosons \cite{GMD} where Alice has a winning and stable strategy.

\subsection{States with different entanglement}
We shall next extend the previous model replacing the CNOT gate with a CU gate given by

\begin{equation}
    CU (|i_1\rangle \otimes |i_0\rangle) = ( U(\theta)^{i_0} |i_1\rangle) \otimes |i_0\rangle,
\end{equation}
where 
\begin{equation}
    U(\theta) = e^{-i\frac{\theta}{2}X} \,  , 
\end{equation}
is a rotation around the $x$-axis of angle $\theta$. 
The rotated Bell operators will be defined as 

\begin{equation}
    RBell(i_{1},i_{0}) = CU(id \otimes H)(X^{i_{1}} \otimes X^{i_{0}}).
    \label{eq51}
\end{equation}

Obviously, for  $\theta=0$ we recover the states and operators considered previously. 

The operators (\ref{eq51}) will be denoted as 

\begin{equation}
\begin{split}
     O_{0}' = RBell(0,0); & \quad O_{1}' = RBell(0,1);\\  O_{2}' = RBell(1,0); & \quad O_{3}' = RBell(1,1).
    \end{split}
    \label{eq53}
\end{equation}
and using them Alice and Bob will  generate the states

\begin{equation}
    |\psi'_{a_{0},b_{0}}\rangle = O^{'A}_{a_{0}}O^{'B}_{b_{0}}|00\rangle \, . 
    \label{eq54new}
\end{equation}

\begin{table}[h!]
\centering
\resizebox{8cm}{!}{
\begin{tabular}{|r|r|r|r|r|r|r|r|r|r|r|r|r|r|r|r|r|}
\hline
 $G'$    & 00 & 22 & 12 & 30 & 13 & 31 & 01 & 23 & 02 & 20 & 10 & 32 & 11 & 33 & 03 & 21 \\ \hline
00 & 1  & 1  & $c$  & $c$  & $s$  & $s$  & 0  & 0  & 0  & 0  & 0  & 0  & 0  & 0  & 0  & 0  \\ \hline
22 & 1  & 1  & $c$  & $c$  & $s$  & $s$  & 0  & 0  & 0  & 0  & 0  & 0  & 0  & 0  & 0  & 0  \\ \hline
12 & $c$  & $c$  & 1  & 1  & 0  & 0  & $s$  & $s$  & 0  & 0  & 0  & 0  & 0  & 0  & 0  & 0  \\ \hline
30 & $c$  & $c$  & 1  & 1  & 0  & 0  & $s$  & $s$  & 0  & 0  & 0  & 0  & 0  & 0  & 0  & 0  \\ \hline
13 & $-s$  & $-s$  & 0  & 0  & 1  & 1  & $-c$  & $-c$  & 0  & 0  & 0  & 0  & 0  & 0  & 0  & 0  \\ \hline
31 & $-s$  & $-s$  & 0  & 0  & 1  & 1  & $-c$  & $-c$  & 0  & 0  & 0  & 0  & 0  & 0  & 0  & 0  \\ \hline
01 & 0  & 0  & $-s$  & $-s$  & $-c$  & $-c$  & 1  & 1  & 0  & 0  & 0  & 0  & 0  & 0  & 0  & 0  \\ \hline
23 & 0  & 0  & $-s$  & $-s$  & $-c$  & $-c$  & 1  & 1  & 0  & 0  & 0  & 0  & 0  & 0  & 0  & 0  \\ \hline
02 & 0  & 0  & 0  & 0  & 0  & 0  & 0  & 0  & 1  & 1  & $c$  & $c$  & $s$  & $s$  & 0  & 0  \\ \hline
20 & 0  & 0  & 0  & 0  & 0  & 0  & 0  & 0  & 1  & 1  & $c$  & $c$  & $s$  & $s$  & 0  & 0  \\ \hline
10 & 0  & 0  & 0  & 0  & 0  & 0  & 0  & 0  & $c$  & $c$  & 1  & 1  & 0  & 0  & $s$  & $s$  \\ \hline
32 & 0  & 0  & 0  & 0  & 0  & 0  & 0  & 0  & $c$  & $c$  & 1  & 1  & 0  & 0  & $s$  & $s$  \\ \hline
11 & 0  & 0  & 0  & 0  & 0  & 0  & 0  & 0  & $-s$ & $-s$  & 0  & 0  & 1  & 1  & $-c$  & $-c$  \\ \hline
33 & 0  & 0  & 0  & 0  & 0  & 0  & 0  & 0  & $-s$ & $-s$  & 0  & 0  & 1  & 1  & $-c$  & $-c$  \\ \hline
03 & 0  & 0  & 0  & 0  & 0  & 0  & 0  & 0  & 0  & 0  & $-s$  & $-s$  & $-c$  & $-c$  & 1  & 1  \\ \hline
21 & 0  & 0  & 0  & 0  & 0  & 0  & 0  & 0  & 0  & 0  & $-s$  & $-s$  & $-c$  & $-c$  & 1  & 1  \\ \hline
\end{tabular}}
 \caption{\small{Metric $G'_{(i_{1},j_{1}),(i_{2},j_{2})}$ as a function of the parameter $\theta$ where $c=\cos\frac{\theta}{2}$ and $s=i\sin\frac{\theta}{2}$.}}
 \label{tabla15}
\end{table}

The entanglement of the states \eqref{eq54new} depends on the value of $\theta$. 
The density matrix obtained by tracing over the qubit 0 reads

\begin{equation}
    \rho'_1 = \frac{1}{2}\begin{pmatrix}
    1 & k\\
    k^* & 1
    \end{pmatrix},
    \quad
    k= \pm i \sin (\theta/2) \ , 
\end{equation}
and has  eigenvalues  $\lambda_\pm = \frac{1}{2}( 1 \pm \sin (\theta/2))$ that are invariant  under the replacement 
 $\theta \rightarrow 2 \pi - \theta$. This symmetry allows us to restrict ourselves to the interval $\theta \in [0, \pi]$. 
The maximal entanglement corresponds to $\theta=0$, and the minimum to $\theta=\pi$ where the states \eqref{eq54new} 
become separable.

The metric $G'_{(i_{1},j_{1}),(i_{2},j_{2})}$ is defined  by 

\begin{equation}
\begin{split}
     G'_{(i_{1},j_{1}),(i_{2},j_{2})} & = \langle 00|O'^{\dagger}_{j_{1}}O'^{\dagger}_{i_{1}}O'_{i_{2}}O'_{j_{2}}|00\rangle , 
    \end{split}
    \label{eq55}
\end{equation}
and its entries are given in Table \ref{tabla15}, which coincide with those in Table \ref{tabla13} for $\theta=0$.
When $\theta \neq 0$ these entries $G'$ can be organized into the following pairs

\barray 
P_1 & = &  \{ 00,22 \}, P_2 = \{ 12,30 \} , \nonumber  \\
P_3 &  = & \{ 13,31 \}, P_4 = \{ 01,23 \} , \label{Pairs}  \\
P_5 & = & \{ 02,20 \}, P_6 = \{ 10,32 \} , \nonumber  \\
P_7 & =  & \{ 11,33 \}, P_8 = \{ 03,21 \}   \, , \nonumber
\earray 
that are related to the sets  \eqref{eq46} as 
\barray 
Set_1 & =  &  P_1 \cup P_2, \quad Set_2  =    P_3 \cup P_4  \, , 
\label{PS} \\
Set_3 & =  &  P_5 \cup P_6, \quad Set_4  =    P_7 \cup P_8 \, .  
\nonumber 
\earray 

We shall next  analyze the different strategies of Alice and Bob. 
Suppose that Alice chooses at random  among the pairs given in eq. (\ref{Pairs}) and that  Bob chooses $b_{0} = 0$.
This leads to the following cases: 

\begin{itemize}
	   
\item Alice selects $(a_1, a_2) \in P_{1}$ then Bob, applying the orthogonality rule \eqref{eq44}, 
	    will choose $(b_1, b_2) \in P_{4,5,6,7,8}$.  The payoffs \eqref{eq45} for $a_0=0,1,2,3$ and $b_0=0$ are given by 
	    
\begin{table}[h!]
\centering	    
\begin{tabular}{|r|r|r|r|r|}
\hline 
$f^A$ & 00 & 10 & 20 &  30 \\ \hline
$P_1$ & 1 & 0 & 0 & $c^2$ \\
\hline
\end{tabular} \quad
\begin{tabular}{|c|c|c|c|c|}
\hline 
$f^B$ & 00 & 10 & 20 &  30 \\ \hline
$P_4$ & 0 & 0 & 0 & $|s|^2$  \\ \hline 
$P_5$ & 0 & $c^2$ & 1 & 0 \\ \hline 
$P_6$ & 0 & 1 & $c^2$ & 0 \\ \hline 
$P_7$ & 0 & 0 & $|s|^2$ & 0 \\ \hline 
$P_8$ & 0 & $|s|^2$  & 0 & 0 \\ \hline 
\end{tabular}
\end{table}

\item Alice selects $(a_1, a_2) \in P_{3}$ then Bob, applying the orthogonality rule \eqref{eq44}, 
	    will choose $(b_1, b_2) \in P_{2,5,6,7,8}$.  The payoffs \eqref{eq45} are given by 
	     
\begin{table}[h!]
\centering	    
\begin{tabular}{|r|r|r|r|r|}
\hline 
$f^A$ & 00 & 10 & 20 &  30 \\ \hline
$P_3$ & $|s|^2$  & 0 & 0 & $0$ \\
\hline
\end{tabular} \quad
\begin{tabular}{|c|c|c|c|c|}
\hline 
$f^B$ & 00 & 10 & 20 &  30 \\ \hline
$P_2$ & $c^2$ & 0 & 0 & $1$  \\ \hline 
$P_5$ & 0 & $c^2$ & 1 & 0 \\ \hline 
$P_6$ & 0 & 1 & $c^2$ & 0 \\ \hline 
$P_7$ & 0 & 0 & $|s|^2$ & 0 \\ \hline 
$P_8$ & 0 & $|s|^2$  & 0 & 0 \\ \hline 
\end{tabular}
\end{table}

\item The cases where Alice selects $P_{2,4,5,6,7,8}$ are evenly distributed among the previous ones. 

\end{itemize}

Taking into account that $1 + c^2 \geq |s|^2$, 
 the best strategy for Bob is  to select the pairs $P_5$ and $P_6$ with equal probability. 
 In this case, the  average payoffs of Alice and Bob are given by
 \barray
 \langle f^A \rangle & = & \frac{1}{2} \left(  \frac{ 1 + c^2}{4} + \frac{ |s|^2}{4} \right) = \frac{ 1}{4} \, , 
 \label{ave} \\
  \langle f^B \rangle & = & \frac{1}{2} \left(  \frac{ 1 + c^2}{4} + \frac{ 1+ c^2}{4} \right) = \frac{ 1+ c^2}{4} \, , 
  \nonumber 
 \earray 
 that yields the normalized probabilities
 \beq
 P_A = \frac{ 1}{2 + c^2}, \qquad P_B = \frac{ 1 + c^2}{ 2 + c^2} \, . 
 \label{Ptot2}
 \eeq
 
\begin{figure}[h!]
    \centering
    \includegraphics[width=8cm]{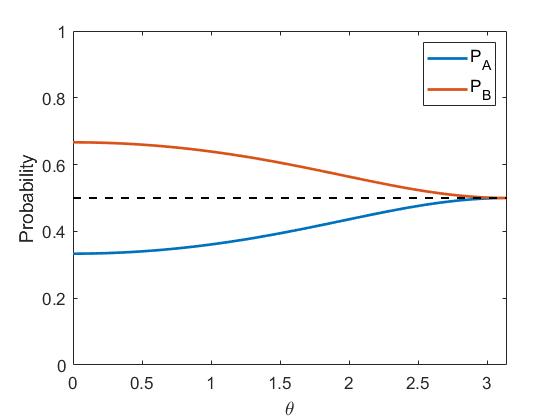}
 \caption{Normalized probabilities  \eqref{Ptot2}  as a function of $\theta \in [0, \pi]$}
  \label{PAB}
\end{figure}

 For $\theta =0$, one recovers eq.  \eqref{Ptot}, while for $\theta = \pi$, one gets  $P_A = P_B = 1/2$. 
 Fig. \ref{PAB} plots  the values of \eqref{Ptot2}, that shows. 
 The highest the entanglement of the states \eqref{eq54new} the highest probability has Bob for  winning over Alice.
 Indeed the ratio of their probabilities can be related to the purity of the states, ${\rm tr} ( {\rho'_1}^2)$, 
 \beq
 \frac{P_B}{P_A} = 3 - 2 {\rm tr} ( {\rho'_1}^2) \ . 
 \label{pu}
 \eeq
 
\subsection{The order matters}

In the quantum games based on the boson, hard-core boson and one-qubit,
the operators $O^A$ and $O^B$,  that generate the joint state,  commute. 
However, in the two-qubit game, the operators \eqref{eq30} or \eqref{eq53} 
do not. This implies that the order of their action on the initial state
may lead to different outcomes. We shall next analyze the case $\theta =0$. 
The joint state is now created by Alice acting first and followed by Bob 
\beq
|\widetilde{\psi}_{a_0, b_0} \rangle = O^B_{b_0} O^A_{a_0} |00 \rangle \, . 
\label{eq60} 
\eeq

The definition of the guess states \eqref{eq39}  remains the same. 
This implies that the overlaps between the guess states and the
joint state  yield a matrix  $\tilde{G}$ that is the partial transposed
of the matrix $G$, that is
\beq
\widetilde{G}_{(i, j), (a_0, b_0)} =  G_{(i, j), (b_0, a_0)}  \, . 
\label{eq61}
\eeq
Hence,  the payoffs on the Alice-first game, are  given by those of the
Bob-first game  as 

\begin{equation}
\begin{split}
     \tilde{f}^A_{(a_1, a_2),(a_0, b_0)}= f^A_{(a_1, a_2),(b_0, a_0)},\\
     \tilde{f}^B_{(b_1, b_2),(a_0, b_0)} =  f^B_{(b_1, b_2),(b_0, a_0)}.
\end{split}
\label{eq62}
\end{equation}

In this new game, one  can  expect  that Alice will have some advantage
because her choice of $a_0$ will restrict possible winning guesses.
Nevertheless, this does not happens because Alice's guess is done before
Bob's one, and then she will reveal information about
the value of $a_0$  to Bob. This leads in turn to a symmetric game. 
Let us illustrate this result with an example. 
Suppose that Alice chooses $a_0=0$. Replacing $b_0$ in  eq.\eqref{bSet}
by $a_0$, one finds that the winning sets are $Set_1$ or $Set_3$. 
So, she will choose at random among both. 
However, Bob, knowing
this information, will notice that $a_0$ is equal to 0 or
2. Then, he will choose the set which is not chosen
by Alice among the pair  of sets.

	    \begin{table}[h!]
\centering	    
\begin{tabular}{|r|r|r|r|r|}
\hline 
$\tilde{f}^A$ & 00 & 01 & 02 &  03 \\ \hline
$Set_1$ & 1 & 0 & 0 & 1 \\
\hline
\end{tabular},  \quad 
\begin{tabular}{|r|r|r|r|r|}
\hline 
$\tilde{f}^B$ & 00 & 01 & 02 &  03 \\ \hline
$Set_3$ & 0 & 1 & 1 & 0 \\
\hline
\end{tabular}
\end{table}

This scenario ends
with each one winning half of the times on average.
The same probabilities will appear for every value of
$a_0$, so the normalized probabilities are 
\beq 
P_A = P_B = \frac{1}{2} \ . 
\eeq

The sets ${\cal C}, {\cal \tilde{C}}$ and the space $ {\cal G}$ involved in the two-qubit game are given by  
\barray 
{\cal C} & =  &  \{ 0,1 \}, \;  {\cal \tilde{C}} =  \{ (0, 0), (0,1), (1,0), (1,1) \} ,\; \label{2q63} \\
 {\cal G} &  =  &  {\rm Span} \{ |00 \rangle, |01 \rangle, |10 \rangle, |11 \rangle \},  
\nonumber 
\earray 
where the distance in ${\cal G}$ is defined by \eqref{b17}.

\section{Simulation on  an  IBM quantum computer}

The previous games are completely characterized by the metric $G$  given in eq.\eqref{eq43},  and its extended version  $G'$ given in \eqref{eq55}. 
Their values are given in Tables \ref{tabla13} and \ref{tabla15} respectively. 
In this section, we shall  present the results obtained using an IBM quantum computer
to find the matrix $G$  that corresponds to the case with $\theta =0$. We have also studied
the case $\theta =\pi$, but it will not be presented here. 

To compute the matrix $G$ we apply two  $O_i$-operators and  two $O^\dagger_j$-operators on  the state $|0 0 \rangle$.
The probability of measuring the state $|0 0\rangle$ provides  $|G|^2$. Some entries of $G$ have phases
but they are  not relevant  to the strategies of the game. Nevertheless, we have included  them in  the results. 

The circuit to compute $G_{(2,2),(3,0)}$ is given by

\begin{figure}[h!]
    \centering
    \includegraphics[width=8cm]{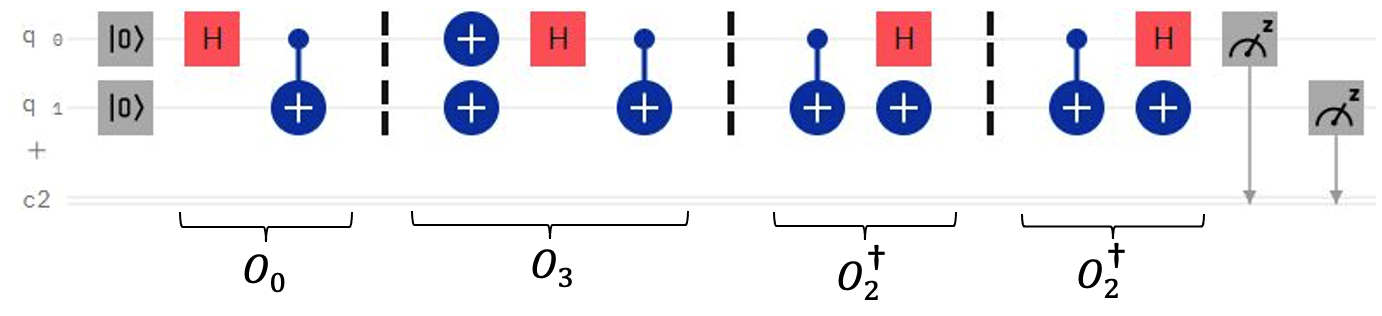}
\end{figure}

We used the qubits $q0$ and $q1$ of the quantum computer ibmq-Manila whose topology is 
\begin{figure}[h!]
    \centering
    \includegraphics[width=6cm]{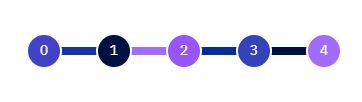}
\end{figure}

\begin{figure}[h!]
    \centering
    \includegraphics[width=6.5cm]{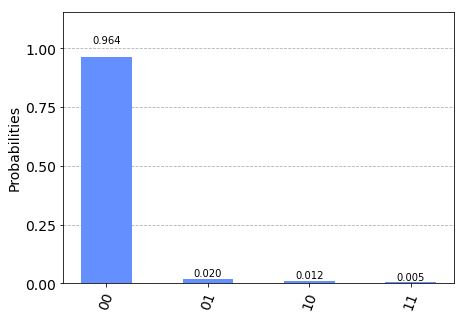}
    \caption{ Histogram of the probabilities obtained to measure $|G_{(2,2),(3,0)}|^2$ after 8192 shots}
    \label{figure5}
\end{figure}

The number of shots was set to its maximum, 8192, obtaining the results plotted in Fig. \ref{figure5}. 
The experimental  value of $0.964$ is very closed to the theoretical value of 1 given in Table \ref{tabla13}. 
This is of course due to the noise and decoherence in this computer.  We have determined experimentally
all the entries of $|G|$ and collect them in Table \ref{tabla16}, including the signs. 
The theoretical values equal to 1 have  an averaged error of 2-3\%, while the null values
have an average error of 10-24\%. This  implies that the orthogonality condition  \eqref{eq41} 
has to a more  relaxed,  say to  $|\langle g_A | g_B \rangle | \lessapprox 0.25$, if the game is to be close
to the theoretical one. 

\vspace{0.5cm} 

	\begin{table}[h!]
	\renewcommand{\arraystretch}{1.2}
\centering
\resizebox{8cm}{!}{
\begin{tabular}{|r|r|r|r|r|r|r|r|r|r|r|r|r|r|r|r|r|}
\hline
  $G_{\rm exp}$ & 00 & 22 & 12 & 30 & 13 & 31 & 01 & 23 & 02 & 20 & 10 & 32 & 11 & 33 & 03 & 21 \\ \hline
00 & 0,97  & 0,98  & 0,97  & 0,97  & 0,24  & 0,24  & 0,20  & 0,24  & 0,20  & 0,20  & 0,22  & 0,20  & 0,10  & 0,10  & 0,17  & 0,10  \\ \hline
22 & 0,98  & 0,97  & 0,97  & 0,98  & 0,24  & 0,24  & 0,24  & 0,24 & 0,20  & 0,20  & 0,20  & 0,22  & 0,14  & 0,10 & 0,10  & 0,10  \\ \hline
12 & 0,97  & 0,97  & 0,97  & 0,97  & 0,24  & 0,24  & 0,24  & 0,24  & 0,20  & 0,22  & 0,20  & 0,20  & 0,10  & 0,10  & 0,14  & 0,14  \\ \hline
30 & 0,97  & 0,97  & 0,98  & 0,98  & 0,24  & 0,24 & 0,22  & 0,24  & 0,20  & 0,20  & 0,24  & 0,22  & 0,10  & 0,10 & 0,14  & 0,10 \\ \hline
13 & 0,26  & 0,24  & 0,24  & 0,22  & 0,97  & 0,97  & -0,97  & -0,97  & 0,10  & 0,10  & 0,10  & 0,10  & 0,20  & 0,20  & 0,22  & 0,20  \\ \hline
31 & 0,24  & 0,24  & 0,22  & 0,24 & 0,95  & 0,97  & -0,97  & -0,97  & 0,17  & 0,10  & 0,10   & 0,10   & 0,22  & 0,20  & 0,22  & 0,20  \\ \hline
01 & 0,26  & 0,24  & 0,24  & 0,22  & -0,97  & -0,97  & 0,96  & 0,98  & 0,10  & 0,14  & 0,10  & 0,10  & 0,20  & 0,22  & 0,22  & 0,20  \\ \hline
23 & 0,24  & 0,24 & 0,24  & 0,24  & -0,96  & -0,95  & 0,97  & 0,98  & 0,14  & 0,10  & 0,10  & 0,14 & 0,20  & 0,20  & 0,20  & 0,22  \\ \hline
02 & 0,22  & 0,20  & 0,22  & 0,20  & 0,14  & 0,14  & 0,10  & 0,10  & 0,98  & 0,97  & 0,97  & 0,97  & 0,26  & 0,24  & 0,24  & 0,24  \\ \hline
20 & 0,22  & 0,24  & 0,24  & 0,22  & 0,14  & 0,10 & 0,10  & 0,10  & 0,97  & 0,97  & 0,97  & 0,97  & 0,24  & 0,24  & 0,26  & 0,24 \\ \hline
10 & 0,20  & 0,20  & 0,22  & 0,20  & 0,10  & 0,14  & 0,10  & 0,10  & 0,97  & 0,97  & 0,97  & 0,97  & 0,22  & 0,24  & 0,28  & 0,24  \\ \hline
32 & 0,20  & 0,20  & 0,22  & 0,20  & 0,10  & 0,10  & 0,10  & 0,10 & 0,95  & 0,97  & 0,95  & 0,97  & 0,24  & 0,22 & 0,24  & 0,24  \\ \hline
11 & 0,10  & 0,10  & 0,10  & 0,14  & 0,22  & 0,20  & 0,20  & 0,20  & 0,24  & 0,24  & 0,24  & 0,24  & 0,98  & 0,97  & -0,95  & -0,97  \\ \hline
33 & 0,10  & 0,10 & 0,10  & 0,10  & 0,20  & 0,20  & 0,20  & 0,22  & 0,24  & 0,22  & 0,26  & 0,22 & 0,97  & 0,97  & -0,97  & -0,97  \\ \hline
03 & 0,10  & 0,10  & 0,10  & 0,10  & 0,20  & 0,20  & 0,20  & 0,20  & 0,24  & 0,24  & 0,26  & 0,22  & -0,97  & -0,97  & 0,95  & 0,97  \\ \hline
21 & 0,10  & 0,10  & 0,10  & 0,10 & 0,20  & 0,22  & 0,20  & 0,20  & 0,26  & 0,24 & 0,24  & 0,24  & -0,97  & -0,96 & 0,96  & 0,97  \\ \hline
\end{tabular}}
\caption{\small{Experimental values of $G_{(i_{1},j_{1}),(i_{2},j_{2})}$ measured obtained on ibmq-Manila.}}
\label{tabla16}
\end{table}

\section{Conclusions and prospects}

We have presented in this work a general formalism of the Chinos game between two players
that includes the quantum games introduced in reference \cite{GMD} using
a boson and extend them to a hard-core boson, a  one-qubit system, and a  two-qubits system. 

A general Chinos game is defined by: i)  two sets, ${\cal C}, \tilde{\cal C}$, containing  the players'  choices and guesses,
ii) a device that maps the previous data to a metric space ${\cal G}$ whose distance is used to compute the players' payoffs,
iii) a restriction rule on the second player,  and iv) an intelligence rule to optimize the players' guesses. In the games
we have considered, the sets ${\cal C}, \tilde{\cal C}$ are a discrete and finite collection of integers. In the classical
Chinos game the space ${\cal G}$ is also discrete and finite, but in the quantum Chinos games ${\cal G}$  is a Hilbert space. 
We have employed the trace norm of density matrices to define the payoffs and the restriction rule, which
for pure states is in direct correspondence with the fidelity. However,  one can consider mixed states in which
case the two  criteria, i.e. distance or fidelity,  can lead to different results. 

The game based on two-qubits shows an interesting interplay between entanglement and the  success probabilities' 
of the  players. Moreover,  the game is sensitive to the way  the joint state is created. In the one and two
qubit games, the device can be implemented on a quantum computer, which has allowed us to simulated the basic matrix
involved in the game using  an IBM quantum computer.  

The rules of the Chinos game make it similar to those considered in the quantum decision theory whose mathematics is based on 
the theory of quantum measurement \cite{Y08}-\cite{Y20}. These classes of games  can also be categorized as {\em abstract economics}  because
the player's strategies are not independent of each other \cite{K18}.  
Finally, we would like to notice that the formalism presented here can be extended in a natural way to more than two players. 
It will be interesting to explore its application to quantum communication protocols.

\vspace{0.5cm} 

\textit{Acknowledgements.-}

 GS acknowledges financial support through the Spanish MINECO grant PGC2018- 095862-B-C21, 
 the Comunidad de Madrid grant  No. S2018/TCS-4342, 
 the Centro de Excelencia Severo Ochoa Program 
 SEV-2016-0597 
 and the CSIC Research Platform on Quantum Technologies PTI-001.

\end{document}